\documentclass{article}
\usepackage[english]{babel}
\usepackage[cp1251]{inputenc}
\usepackage{amsmath,amssymb}
\begin{document}
\begin{center}
\textbf{\Large{The Dirichlet problem for the generalized bi-axially symmetric Helmholtz equation}}\\
\medskip
\textbf{M.S. Salakhitdinov, A. Hasanov}\\
\medskip
Communicated by Sh.A.Alimov
\end{center}
\medskip



\textbf{Abstract.} In [18], fundamental solutions for the generalized bi-axially symmetric Helmholtz equation were constructed in $R_2^ +   = \left\{ {\left( {x,y} \right):x > 0,y > 0} \right\}.$ They contain Kummer's confluent hypergeometric functions in three variables. In this paper, using one of the constructed fundamental solutions, the Dirichlet problem is solved in the domain $\Omega  \subset R_2^ +.$ Using the method of Green's functions, solution of this problem is found  in an explicit form.\\

\section{Introduction.}

In the monograph of Gilbert [16], by applying methods of complex analysis, integral representations of solutions of the generalized bi-axially Helmholtz equation
$$
H_{\alpha ,\beta }^\lambda  \left( u \right) \equiv u_{xx}  + u_{yy}  + \frac{{2\alpha }}
{x}u_x  + \frac{{2\beta }} {y}u_y  - \lambda ^2 u = 0, \eqno \left( H_{\alpha ,\beta }^\lambda \right)
$$
were constructed via analytic functions. Here $0 < 2\alpha ,2\beta  < 1,$ $\alpha ,\beta ,\lambda $ are constants. When $
\lambda  = 0 $ this equation is known as the equation of generalized axially symmetric potential theory. This terminology was used for the first time by Weinstein, who first considered fractional dimensional spaces in the potential theory [33, 34]. The special case whith $\,\lambda  = 0 $
has also been investigated by Erdelyi [5, 6], Gilbert [9-15], Ranger [29], Henrici [21, 22]. There are many works [1-3, 8, 17, 23, 25-28, 30, 32] in which some problems for equation $\left(H_{\alpha ,\beta }^\lambda\right)$ were studied. In the paper [18] for equation $\left(H_{\alpha ,\beta }^\lambda\right)$ the following fundamental solutions on $ R_2^ +   = \left\{ {\left( {x,y} \right):x > 0,y > 0} \right\}:$ have been constructed
$$
q_1 \left( {x,y;x_0 ,y_0 } \right) = k_1 \left( {r^2 } \right)^{ - \alpha  - \beta } A_2^{\left( 3 \right)} \left( {\alpha  + \beta ;\alpha ,\beta ;2\alpha ,2\beta ;\xi ,\eta ,\zeta } \right),\eqno(1.1)
$$
$$
q_2 \left( {x,y;x_0 ,y_0 } \right) = k_2 \left( {r^2 } \right)^{\alpha  - \beta  - 1} x^{1 - 2\alpha } x_0^{1 - 2\alpha } A_2^{\left( 3 \right)} \left( {1 - \alpha  + \beta ;1 - \alpha ,\beta ;2 - 2\alpha ,2\beta ;\xi ,\eta ,\zeta } \right),\eqno(1.2)
$$
$$
q_3 \left( {x,y;x_0 ,y_0 } \right) = k_3 \left( {r^2 } \right)^{ - \alpha  + \beta  - 1} y^{1 - 2\beta } y_0^{1 - 2\beta } A_2^{\left( 3 \right)} \left( {1 + \alpha  - \beta ;\alpha ,1 - \beta ;2\alpha ,2 - 2\beta ;\xi ,\eta ,\zeta } \right),\eqno(1.3)
$$
$$
\begin{array}{l}
  q_4 \left( {x,y;x_0 ,y_0 } \right) \hfill \\
   = k_4 \left( {r^2 } \right)^{\alpha  + \beta  - 2} x^{1 - 2\alpha } y^{1 - 2\beta } x_0^{1 - 2\alpha } y_0^{1 - 2\beta } A_2^{\left( 3 \right)} \left( {2 - \alpha  - \beta ;1 - \alpha ,1 - \beta ;2 - 2\alpha ,2 - 2\beta ;\xi ,\eta ,\zeta } \right), \hfill \\
\end{array}\eqno(1.4)
$$
where
$$
k_1  = \frac{{2^{2\alpha  + 2\beta } }}
{{4\pi }}\frac{{\Gamma \left( \alpha  \right)\Gamma \left( \beta  \right)\Gamma \left( {\alpha  + \beta } \right)}}
{{\Gamma \left( {2\alpha } \right)\Gamma \left( {2\beta } \right)}},\eqno(1.5)
$$
$$
k_2  = \frac{{2^{2 - 2\alpha  + 2\beta } }}
{{4\pi }}\frac{{\Gamma \left( {1 - \alpha } \right)\Gamma \left( \beta  \right)\Gamma \left( {1 - \alpha  + \beta } \right)}}
{{\Gamma \left( {2 - 2\alpha } \right)\Gamma \left( {2\beta } \right)}},\eqno(1.6)
$$
$$
k_3  = \frac{{2^{2 + 2\alpha  - 2\beta } }}
{{4\pi }}\frac{{\Gamma \left( \alpha  \right)\Gamma \left( {1 - \beta } \right)\Gamma \left( {1 + \alpha  - \beta } \right)}}
{{\Gamma \left( {2\alpha } \right)\Gamma \left( {2 - 2\beta } \right)}},\eqno(1.7)
$$
$$
k_4  = \frac{{2^{4 - 2\alpha  - 2\beta } }}
{{4\pi }}\frac{{\Gamma \left( {1 - \alpha } \right)\Gamma \left( {1 - \beta } \right)\Gamma \left( {2 - \alpha  - \beta } \right)}}
{{\Gamma \left( {2 - 2\alpha } \right)\Gamma \left( {2 - 2\beta } \right)}},\eqno(1.8)
$$
$$
r^2=\left(x-x_0\right)^2+\left(y-y_0\right)^2,\,\,r_1^2=\left(x+x_0\right)^2+\left(y-y_0\right)^2,\,\,r_2^2=\left(x-x_0\right)^2+\left(y+y_0\right)^2,
$$
$$
\xi  = \frac{{r^2  - r_1^2 }}
{{r^2 }},
\eta  = \frac{{r^2  - r_2^2 }}
{{r^2 }},
\zeta  =  - \frac{{\lambda ^2 }}
{4}r^2 ,\eqno(1.9)
$$
$$
A_2^{\left( 3 \right)} \left( {a;b_1 ,b_2 ;c_1 ,c_2 ;x,y,z} \right) = \sum\limits_{m,n,p = 0}^\infty  {} \frac{{\left( a \right)_{m + n - p} \left( {b_1 } \right)_m \left( {b_2 } \right)_n }}
{{\left( {c_1 } \right)_m \left( {c_2 } \right)_n m!n!p!}}x^m y^n z^p ,\eqno(1.10)
$$
and $\left( a \right)_n  = \Gamma \left( {a + n} \right)/\Gamma \left( a \right)$ is the Pochhammer symbol.

\section{Green's formulas.}

We consider an identity
$$
x^{2\alpha } y^{2\beta } \left[ {uH_{\alpha ,\beta }^\lambda  \left( v \right) - vH_{\alpha ,\beta }^\lambda  \left( u \right)} \right] = \frac{\partial }
{{\partial x}}\left[ {x^{2\alpha } y^{2\beta } \left( {v_x u - vu_x } \right)} \right] + \frac{\partial }
{{\partial y}}\left[ {x^{2\alpha } y^{2\beta } \left( {v_y u - vu_y } \right)} \right].\eqno(2.1)
$$
Integrating both parts of identity (2.1) over $\Omega  \subset R_2^ +$, and using Green's formula we find
$$
\int\limits_\Omega  {}x^{2\alpha } y^{2\beta } \left[ {uH_{\alpha ,\beta }^\lambda  \left( v \right) - vH_{\alpha ,\beta }^\lambda  \left( u \right)} \right]dxdy = \int\limits_S {} x^{2\alpha } y^{2\beta } u\left( {v_x dy - v_y dx} \right) - x^{2\alpha } y^{2\beta } v\left( {u_x dy - u_y dx} \right),\eqno(2.2)
$$
where $S = \partial \Omega$ is the boundary of the domain $\Omega.$ Formula (2.2) named as Green's formula is deduced under the following assumptions: \\
- the functions $u$ and $v$ are continuous on the closure of the domain $\Omega$, i.e. on $\bar \Omega,$ \\
- the partial derivatives of the first and second orders of $u$ and $v$ are continuous on $\Omega,$ \\
- the integrals over $\Omega,$ containing partial derivatives of the first and second orders of $u$ and $v$ have sense. \\
If  $u$, $v$ are solutions of the equation $\left(H_{\alpha ,\beta }^\lambda\right)$, then by formula (2.2) we get
$$
\int\limits_S {} x^{2\alpha } y^{2\beta } \left( {u\frac{{\partial v}}
{{\partial n}} - v\frac{{\partial u}}
{{\partial n}}} \right)ds = 0,\eqno(2.3)
$$
where
$$
\frac{\partial }
{{\partial n}} = \frac{{dy}}
{{ds}}\frac{\partial }
{{\partial x}} - \frac{{dx}}
{{ds}}\frac{\partial }
{{\partial y}}, \,\,\frac{{dy}}
{{ds}} = \cos \left( {n,x} \right), \,\,\, \frac{{dx}}
{{ds}} =  - \cos \left( {n,y} \right),\eqno(2.4)
$$
$n$ is the exterior normal to the curve $S.$ The following identity also takes place:
$$
\int\limits_\Omega  {}x^{2\alpha } y^{2\beta } \left[ {u_x^2  + u_y^2  + \lambda ^2 u^2 } \right]dxdy = \int\limits_S {} x^{2\alpha } y^{2\beta } u\frac{{\partial u}}
{{\partial n}}ds,\eqno(2.5)
$$
where $u$ is a solution of the equation $\left(H_{\alpha ,\beta }^\lambda\right)$.

\section{The formulation and the uniqueness of the Dirichlet problem. }

Let $\Omega  \subset R_2^ +   = \left\{ {\left( {x,y} \right):x > 0,y > 0} \right\}$ be a domain limited by intervals $I_1  = \left( {0,a} \right),a = const > 0$, $I_2  = \left( {0,b} \right),b = const > 0$ of the axis $OX$, $OY$ respectively, and a curve $\Gamma $ with endpoints $A\left( {a,0} \right),$ $B\left( {0,b} \right).$ The parametrical equation of the curve $\Gamma $ will be $x = x\left( s \right), \,y = y\left( s \right),$ where $s$ is the length of the arc counted from the point $A\left( {a,0} \right).$ Concerning the curve $\Gamma $ we shall assume that: \\
- the functions $x = x\left( s \right),y = y\left( s \right)$ have continuous derivatives $x'\left( s \right),y'\left( s \right)$
 on the segment $\left[ {0,l} \right],$ not simultaneously equal to zero, where $l$ is length of the curve $\Gamma$; \\
- the derivatives $x''\left( s \right),y''\left( s \right)$ satisfy to the H\"{o}lder condition on $\left[ {0,l} \right]$;  \\
- in neighborhoods of the points $A\left( {a,0} \right)$ and $B\left( {0,b} \right)$ the conditions
$$
\left| {\frac{{dx}} {{ds}}} \right| \leq Cy^{1 + \varepsilon } \left( s \right), \,\,
\left| {\frac{{dy}} {{ds}}} \right| \leq Cx^{1 + \varepsilon } \left( s \right), \,\, 0 < \varepsilon  < 1, \,\, C = const,
$$
are satisfied. \\
\textbf{Problem $D.$} Find a solution $u$ of equation $\left(H_{\alpha ,\beta }^\lambda\right)$ belonging to the class $C\left( {\bar \Omega } \right) \cap C^2 \left( {\Omega } \right),$ satisfying the conditions
$$
\left. {u\left( {x,y} \right)} \right|_{y = 0}  = \tau _1 \left( x \right),\,\, x \in \bar I_1 ,\eqno(3.1)
$$
$$
\left. {u\left( {x,y} \right)} \right|_{x = 0}  = \tau _2 \left( y \right), \,\, y \in \bar I_2 ,\eqno(3.2)
$$
$$
\left. {u\left( {x,y} \right)} \right|_\Gamma   = \varphi \left( s \right),\,\,  0 \leq s \leq l,\eqno(3.3)
$$
where $\tau _1,\tau _2,\varphi $ are given continuous functions and  $\tau _1 \left( 0 \right) = \tau _2 \left( 0 \right),$ $\tau _1 \left( a \right) = \varphi \left( 0 \right),$ $\tau _2 \left( b \right) = \varphi \left( l \right).$

\textbf{Theorem 1.} If the problem $D$ has a solution in the domain $\Omega $, then it is unique.\\
\textbf{Proof.}  Let $\tau _1 \left( x \right) = \tau _2 \left( y \right) = \varphi \left( s \right) = 0$ then by virtue of identity (2.5), we have
$$
\int\limits_{\Omega } {}x^{2\alpha } y^{2\beta } \left[ {u_x^2  + u_y^2  + \lambda ^2 u^2 } \right]dxdy = 0.\eqno(3.4)
$$
By (3.4) it follows that $u_x \left( {x,y} \right) = u_y \left( {x,y} \right) = u\left( {x,y} \right) = 0.$ Hence, we have $u\left( {x,y} \right) \equiv 0$ in the domain $\Omega $. $\Box$

We note that the uniqueness of a solution of the problem $D$
in the domain $\Omega $ also follows by the extremum principle for elliptic differential equations.

\section{The existence theorem}

Let $a = b$ and $\Gamma =:\{(x,y)\in R_2^+: x^2  + y^2  = a^2\}$. We denote this domain by $\Omega _0$. The function $G_4 \left( {x,y;x_0 ,y_0 } \right)$ satisfying the following conditions is called as Green's function of the problem $D:$ \\
- inside the domain $\Omega _0 ,$ except for the point $\left( {x_0 ,y_0 } \right),$ this function is a regular solution of equation $\left(H_{\alpha ,\beta }^\lambda\right)$; \\
- it satisfies the boundary condition
$$
\left. {G_4 \left( {x,y;x_0 ,y_0 } \right)} \right|_{\Gamma  \cup I_1  \cup I_2 }  = 0;\eqno(4.1)
$$
- it can be represented in the form
$$
G_4 \left( {x,y;x_0 ,y_0 } \right) = q_4 \left( {x,y;x_0 ,y_0 } \right) - \left( {R_0^2 } \right)^{ - \alpha  - \beta } q_4 \left( {x,y;\bar x_0 ,\bar y_0 } \right),\eqno(4.2)
$$
where
$$
R_0^2  = x_0^2  + y_0^2 , \,\, \bar x_0  = \frac{{x_0 }} {{R_0^2 }},\,\, \bar y_0  = \frac{{y_0 }} {{R_0^2 }},\eqno(4.3)
$$
$q_4 \left( {x,y;x_0 ,y_0 } \right)$ is a fundamental solution, $q_4 \left( {x,y;\bar x_0 ,\bar y_0 } \right)$ is a regular solution of equation  $\left(H_{\alpha ,\beta }^\lambda\right)$ in the domain $\Omega _0 .$

Let $\left( {x_0 ,y_0 } \right) \in \Omega _0.$ We cut out from $\Omega _0 $ a circle of small radius $\rho $ with the center at the point $\left( {x_0 ,y_0 } \right)$ and the remaining part of $\Omega _0,$ we denote by $\Omega _0^\rho.$ $C_\rho  $ is a boundary of the cutted out circle. Applying  formula (2.3), we obtain
$$
\begin{array}{l}
\displaystyle \int\limits_{C_\rho  } {x^{2\alpha } y^{2\beta } u\frac{{\partial G_4 \left( {x,y;x_0 ,y_0 } \right)}}
{{\partial n}}ds}  - \int\limits_{C_\rho  } {x^{2\alpha } y^{2\beta } G_4 \left( {x,y;x_0 ,y_0 } \right)\frac{{\partial u}}
{{\partial n}}ds}  \hfill \\
= \displaystyle \int\limits_0^a {\left. {x^{2\alpha } y^{2\beta } \tau _1 \left( x \right)\frac{\partial }
{{\partial y}}G_4 \left( {x,y;x_0 ,y_0 } \right)} \right|_{y = 0} dx}  \hfill \\
+ \displaystyle \int\limits_0^a {x^{2\alpha } y^{2\beta } \tau _2 \left( y \right)\left. {\frac{\partial }
{{\partial x}}G_4 \left( {x,y;x_0 ,y_0 } \right)} \right|_{x = 0} dy}  - \int\limits_\Gamma  {x^{2\alpha } y^{2\beta } \varphi \left( s \right)\frac{{\partial G_4 \left( {x,y;x_0 ,y_0 } \right)}}
{{\partial n}}ds} . \hfill \\
\end{array}\eqno(4.4)
$$
Using the derivation formula
$$
\begin{array}{l}
\displaystyle  \frac{{\partial ^{i + j + k} }}
{{\partial x^i \partial y^j \partial z^k }}A_2^{\left( 3 \right)} \left( {a;b_1 ,b_2 ;c_1 ,c_2 ;x,y,z} \right) \hfill \\
   = \displaystyle \frac{{\left( a \right)_{i + j - k} \left( {b_1 } \right)_i \left( {b_2 } \right)_j }}
{{\left( {c_1 } \right)_i \left( {c_2 } \right)_j }}A_2^{\left( 3 \right)} \left( {a + i + j - k;b_1  + i,b_2  + j;c_1  + i,c_2  + j;x,y,z} \right), \hfill \\
\end{array}\eqno(4.5)
$$
and considering the adjacent relation
$$
\begin{array}{l}
\displaystyle \frac{{ab_1 }}
{{c_1 }}xA_2^{\left( 3 \right)} \left( {1 + a;1 + b_1 ,b_2 ;1 + c_1 ,c_2 ;x,y,z} \right) + \frac{{ab_2 }}
{{c_2 }}yA_2^{\left( 3 \right)} \left( {1 + a;b_1 ,1 + b_2 ;c_1 ,1 + c_2 ;x,y,z} \right) -  \hfill \\
- \displaystyle \frac{1}
{{a - 1}}zA_2^{\left( 3 \right)} \left( {a - 1;b_1 ,b_2 ;c_1 ,c_2 ;x,y,z} \right) \hfill \\
= \displaystyle aA_2^{\left( 3 \right)} \left( {1 + a;b_1 ,b_2 ;c_1 ,c_2 ;x,y,z} \right) - aA_2^{\left( 3 \right)} \left( {a;b_1 ,b_2 ;c_1 ,c_2 ;x,y,z} \right), \hfill \\
\end{array}\eqno(4.6)
$$
we find that
$$
\begin{array}{l}
x^{2\alpha } \displaystyle \frac{\partial }
{{\partial x}}q_4 \left( {x,y;x_0 ,y_0 } \right) \hfill \\
= k_4 \left( {1 - 2\alpha } \right)\left( {r^2 } \right)^{\alpha  + \beta  - 2} x_0^{1 - 2\alpha } \left( {yy_0 } \right)^{1 - 2\beta } A_2^{\left( 3 \right)} \left( {2 - \alpha  - \beta ;1 - \alpha ,1 - \beta ;2 - 2\alpha ,2 - 2\beta ;\xi ,\eta ,\zeta } \right) \hfill \\
- 2k_4 \left( {2 - \alpha  - \beta } \right)x\left( {r^2 } \right)^{\alpha  + \beta  - 3} x_0^{2 - 2\alpha } \left( {yy_0 } \right)^{1 - 2\beta }  A_2^{\left( 3 \right)} \left( {3 - \alpha  - \beta ;2 - \alpha ,1 - \beta ;3 - 2\alpha ,2 - 2\beta ;\xi ,\eta ,\zeta } \right) \hfill \\
- 2k_4 \left( {2 - \alpha  - \beta } \right)x\left( {r^2 } \right)^{\alpha  + \beta  - 3} x_0^{1 - 2\alpha } \left( {yy_0 } \right)^{1 - 2\beta } \left( {x - x_0 } \right) A_2 \left( {3 - \alpha  - \beta ;1 - \alpha ,1 - \beta ;2 - 2\alpha ,2 - 2\beta ;\xi ,\eta ,\zeta } \right) \hfill \\
\end{array}\eqno(4.7)
$$
and
$$
\begin{array}{l}
y^{2\beta } \displaystyle \frac{\partial }
{{\partial y}}q_4 \left( {x,y;x_0 ,y_0 } \right) \hfill \\
= k_4 \left( {1 - 2\beta } \right)\left( {r^2 } \right)^{\alpha  + \beta  - 2} \left( {xx_0 } \right)^{1 - 2\alpha } y_0^{1 - 2\beta } A_2^{\left( 3 \right)} \left( {2 - \alpha  - \beta ;1 - \alpha ,1 - \beta ;2 - 2\alpha ,2 - 2\beta ;\xi ,\eta ,\zeta } \right) \hfill \\
- 2k_4 \left( {2 - \alpha  - \beta } \right)y\left( {r^2 } \right)^{\alpha  + \beta  - 3} \left( {xx_0 } \right)^{1 - 2\alpha } y_0^{2 - 2\beta } A_2^{\left( 3 \right)} \left( {3 - \alpha  - \beta ;1 - \alpha ,2 - \beta ;2 - 2\alpha ,3 - 2\beta ;\xi ,\eta ,\zeta } \right) \hfill \\
- 2k_4 \left( {2 - \alpha  - \beta } \right)y\left( {r^2 } \right)^{\alpha  + \beta  - 3} \left( {xx_0 } \right)^{1 - 2\alpha } y_0^{1 - 2\beta } \left( {y - y_0 } \right) \hfill A_2^{\left( 3 \right)} \left( {3 - \alpha  - \beta ;1 - \alpha ,1 - \beta ;2 - 2\alpha ,2 - 2\beta ;\xi ,\eta ,\zeta } \right). \hfill \\
\end{array}\eqno(4.8)
$$
It is easy to prove that the following formulas are true:
$$
A_2^{\left( 3 \right)} \left( {a;b_1 ,b_2 ;c_1 ,c_2 ;0,y,z} \right) = \sum\limits_{n,p = 0}^\infty  {} \frac{{\left( a \right)_{n - p} \left( {b_2 } \right)_n }}
{{\left( {c_2 } \right)_n n!p!}}y^n z^p  = H_3 \left( {a,b_2 ;c_2 ;y,z} \right),\eqno(4.9)
$$
$$
A_2 \left( {a;b_1 ,b_2 ;c_1 ,c_2 ;x,0,z} \right) = \sum\limits_{m,p = 0}^\infty  {} \frac{{\left( a \right)_{m - p} \left( {b_1 } \right)_m }}
{{\left( {c_1 } \right)_m m!p!}}x^m z^p  = H_3 \left( {a,b_1 ;c_1 ;x,z} \right),\eqno(4.10)
$$
where $H_3 \left( {a,b;c;x,y} \right)$ is Kummer's hypergeometric function in two arguments ([7], p. 221, formula (31)). By virtue of equalities (4.7),(4.8),(4.9), (4.10) and taking into account that $\left. \xi  \right|_{x = 0}  = 0,$ $\left. \eta  \right|_{y = 0}  = 0$, we get
$$
\begin{array}{l}
  y^{2\beta } \left. { \displaystyle  \frac{\partial }
{{\partial y}}G_4 \left( {x,y;x_0 ,y_0 } \right)} \right|_{y = 0}  = k_4 \left( {1 - 2\beta } \right)x_0^{1 - 2\alpha } y_0^{1 - 2\beta } x^{1 - 2\alpha }  \hfill \\
   \times  \displaystyle \left\{ {\frac{{H_3 \left( {2 - \alpha  - \beta ,1 - \alpha ;2 - 2\alpha ;\rho _1 ,\rho _1^* } \right)}}
{{\left[ {\left( {x - x_0 } \right)^2  + y_0^2 } \right]^{2 - \alpha  - \beta } }} - \displaystyle \frac{{H_3 \left( {2 - \alpha  - \beta ,1 - \alpha ;2 - 2\alpha ;\rho _2 ,\rho _2^* } \right)}}
{{\left[ {\left( {a - \displaystyle \frac{{xx_0 }}
{a}} \right)^2  + \displaystyle \frac{1}
{{a^2 }}x^2 y_0^2 } \right]^{2 - \alpha  - \beta } }}} \right\} \hfill \\
\end{array}\eqno(4.11)
$$
and
$$
\begin{array}{l}
  x^{2\alpha } \left. {\displaystyle \frac{\partial }
{{\partial x}}G_4 \left( {x,y;x_0 ,y_0 } \right)} \right|_{x = 0}  = k_4 \left( {1 - 2\alpha } \right)x_0^{1 - 2\alpha } y_0^{1 - 2\beta } y^{1 - 2\beta }  \hfill \\
  \times \displaystyle \left\{ {\frac{{H_3 \left( {2 - \alpha  - \beta ,1 - \beta ;2 - 2\beta ;\rho _3 ,\rho _3^* } \right)}}
{{\left[ {x_0^2  + \left( {y - y_0^2 } \right)^2 } \right]^{2 - \alpha  - \beta } }} - \frac{{H_3 \left( {2 - \alpha  - \beta ,1 - \beta ;2 - 2\beta ;\rho _4 ,\rho _4^* } \right)}}
{{\left[ {\left( {a - \displaystyle \frac{{yy_0 }}
{a}} \right)^2  + \displaystyle \frac{1}
{{a^2 }}x_0^2 y^2 } \right]^{2 - \alpha  - \beta } }}} \right\}, \hfill \\
\end{array}\eqno(4.12)
$$
where
$$
\begin{array}{l}
 \displaystyle \rho _1  = \frac{{ - 4xx_0 }}
{{\left( {x - x_0 } \right)^2  + y_0^2 }},\,\,\,\,\,\,\,\,\,\,\,\,\,\,\,\,\,\,\,\rho _1^*  =  - \frac{{\lambda ^2 }}
{4}\left[ {\left( {x - x_0 } \right)^2  + y_0^2 } \right], \hfill \\
\displaystyle  \rho _2  = \frac{{ - 4xx_0 }}
{{\left( {a - \displaystyle \frac{{xx_0 }}
{a}} \right)^2  + \displaystyle \frac{1}
{{a^2 }}x^2 y_0^2 }},\,\,\rho _2^*  =  - \frac{{a^2 \lambda ^2 }}
{{4R_0^2 }}\left[ {\left( {a - \frac{{xx_0 }}
{a}} \right)^2  + \frac{1}
{{a^2 }}x^2 y_0^2 } \right], \hfill \\
\end{array}
$$

$$
\begin{array}{l}
 \displaystyle \rho _3  = \frac{{ - 4yy_0 }}
{{x_0^2  + \left( {y - y_0 } \right)^2 }},\,\,\,\,\,\,\,\,\,\,\,\,\,\,\,\,\,\,\,\,\rho _3^*  =  - \frac{{\lambda ^2 }}
{4}\left[ {x_0^2  + \left( {y - y_0 } \right)^2 } \right], \hfill \\
\displaystyle  \rho _4  = \displaystyle \frac{{ - 4yy_0 }}
{{\left( {a - \displaystyle \frac{{yy_0 }}
{a}} \right)^2  + \displaystyle \frac{1}
{{a^2 }}x_0^2 y^2 }},\,\,\,\rho _4^*  =  - \displaystyle \frac{{a^2 \lambda ^2 }}
{{4R_0^2 }}\left[ {\left( {a - \displaystyle \frac{{yy_0 }}
{a}} \right)^2  + \displaystyle \frac{1}
{{a^2 }}x_0^2 y^2 } \right]. \hfill \\
\end{array}
$$
Now we shall consider the right-hand side of identity (4.4). Taking into account (4.7) and (4.8), we find that
$$
\begin{array}{l}
\displaystyle   \frac{{\partial q_4 }}
{{\partial n}} =  - k_4 \left( {2 - \alpha  - \beta } \right)\left( {r^2 } \right)^{\alpha  + \beta  - 2} \left( {xx_0 } \right)^{1 - 2\alpha } \left( {yy_0 } \right)^{1 - 2\beta }  \hfill \\
\times A_2^{\left( 3 \right)} \left( {3 - \alpha  - \beta ;1 - \alpha ,1 - \beta ;2 - 2\alpha ,2 - 2\beta ;\xi ,\eta ,\zeta } \right)\displaystyle \frac{\partial }
{{\partial n}}\left[ {\ln \,r^2 } \right] \hfill \\
+ \displaystyle k_4 \left( {r^2 } \right)^{\alpha  + \beta  - 2} x_0^{1 - 2\alpha } y_0^{1 - 2\beta } x^{ - 2\alpha } y^{ - 2\beta } \left[ {\left( {1 - 2\alpha } \right)y\frac{{dy}}
{{ds}} - \left( {1 - 2\beta } \right)x\frac{{dx}}
{{ds}}} \right] \hfill \\
\times A_2^{\left( 3 \right)} \left( {2 - \alpha  - \beta ;1 - \alpha ,1 - \beta ;2 - 2\alpha ,2 - 2\beta ;\xi ,\eta ,\zeta } \right) \hfill \\
- \displaystyle 2k_4 \left( {2 - \alpha  - \beta } \right)\left( {r^2 } \right)^{\alpha  + \beta  - 3} \left( {xx_0 } \right)^{1 - 2\alpha } \left( {yy_0 } \right)^{1 - 2\beta } A_2^{\left( 3 \right)} \left( {3 - \alpha  - \beta ;2 - \alpha ,1 - \beta ;3 - 2\alpha ,2 - 2\beta ;\xi ,\eta ,\zeta } \right)\frac{{dy}}
{{ds}} \hfill \\
+ \displaystyle 2k_4 \left( {2 - \alpha  - \beta } \right)\left( {r^2 } \right)^{\alpha  + \beta  - 3} \left( {xx_0 } \right)^{1 - 2\alpha } \left( {yy_0 } \right)^{1 - 2\beta } A_2^{\left( 3 \right)} \left( {3 - \alpha  - \beta ;1 - \alpha ,2 - \beta ;2 - 2\alpha ,3 - 2\beta ;\xi ,\eta ,\zeta } \right)\frac{{dx}}
{{ds}}. \hfill \\
\end{array}\eqno(4.13)
$$
Further we have
$$
\begin{array}{l}
 \displaystyle  \int\limits_{C_\rho  } {x^{2\alpha } y^{2\beta } u\frac{{\partial G_4 \left( {x,y;x_0 ,y_0 } \right)}}
{{\partial n}}ds}  \hfill \\
   = \displaystyle \int\limits_{C_\rho  } {x^{2\alpha } y^{2\beta } u\frac{{\partial q_4 \left( {x,y;x_0 ,y_0 } \right)}}
{{\partial n}}ds}  - \left( {R_0^2 } \right)^{ - \alpha  - \beta } \int\limits_{C_\rho  } {x^{2\alpha } y^{2\beta } u\frac{{\partial q_4 \left( {x,y;\bar x_0 ,\bar y_0 } \right)}}
{{\partial n}}ds}  = J_1  + J_2 . \hfill \\
\end{array}\eqno(4.14)
$$
Substituting (4.13) in (4.14) and passing to the polar coordinates $x\, = x_0  + \rho \cos \,\varphi ,\,\,y = y_0  + \rho \sin \,\varphi,$ we have
$$
\begin{array}{l}
  \displaystyle   J_1  = 2k_4 \left( {2 - \alpha  - \beta } \right)x_0^{1 - 2\alpha } y_0^{1 - 2\beta } \int\limits_0^{2\pi } {} \left( {x_0  + \rho \cos \,\varphi } \right)\left( {y_0  + \rho \sin \,\varphi } \right)u\left( {x_0  + \rho \cos \,\varphi ,y_0  + \rho \sin \,\varphi } \right) \hfill \\
   \times \left( {\rho ^2 } \right)^{\alpha  + \beta  - 2} A_2^{\left( 3 \right)} \left( {3 - \alpha  - \beta ;1 - \alpha ,1 - \beta ;2 - 2\alpha ,2 - 2\beta ;\xi ,\eta ,\zeta } \right)d\varphi  \hfill \\
   +  \displaystyle  k_4 x_0^{1 - 2\alpha } y_0^{1 - 2\beta } \int\limits_0^{2\pi } {} u\left( {x_0  + \rho \cos \,\varphi ,y_0  + \rho \sin \,\varphi } \right) \hfill \\
   \times  \displaystyle  \left[ {\left( {1 - 2\alpha } \right)y_0 \cos \,\varphi  + \left( {1 - 2\beta } \right)x_0 \sin \varphi  + \left( {1 - \alpha  - \beta } \right)\rho \sin 2\varphi } \right] \hfill \\
  \times  \displaystyle  \left( {\rho ^2 } \right)^{\alpha  + \beta  - 1} A_2^{\left( 3 \right)} \left( {2 - \alpha  - \beta ;1 - \alpha ,1 - \beta ;2 -  \displaystyle  2\alpha ,2 - 2\beta ;\xi ,\eta ,\zeta } \right)d\varphi  \hfill \\
   -  \displaystyle  2k_4 \left( {2 - \alpha  - \beta } \right)x_0^{1 - 2\alpha } y_0^{1 - 2\beta } \int\limits_0^{2\pi } {} \left( {x_0  + \rho \cos \,\varphi } \right)\left( {y_0  + \rho \sin \,\varphi } \right)u\left( {x_0  + \rho \cos \,\varphi ,y_0  + \rho \sin \,\varphi } \right) \hfill \\
  \times \left( {\rho ^2 } \right)^{\alpha  + \beta  - 2} A_2^{\left( 3 \right)} \left( {3 - \alpha  - \beta ;2 - \alpha ,1 - \beta ;3 - 2\alpha ,2 - 2\beta ;\xi ,\eta ,\zeta } \right)\cos \,\varphi d\varphi  \hfill \\
   -  \displaystyle  2k_4 \left( {2 - \alpha  - \beta } \right)x_0^{1 - 2\alpha } y_0^{1 - 2\beta } \int\limits_0^{2\pi } {} \left( {x_0  + \rho \cos \,\varphi } \right)\left( {y_0  + \rho \sin \,\varphi } \right)u\left( {x_0  + \rho \cos \,\varphi ,y_0  + \rho \sin \,\varphi } \right) \hfill \\
   \times \left( {\rho ^2 } \right)^{\alpha  + \beta  - 2} A_2^{\left( 3 \right)} \left( {3 - \alpha  - \beta ;1 - \alpha ,2 - \beta ;2 - 2\alpha ,3 - 2\beta ;\xi ,\eta ,\zeta } \right)\sin \varphi d\varphi  \hfill \\
   = J_{11}  + J_{12}  + J_{13}  + J_{14} . \hfill \\
\end{array}\eqno(4.15)
$$
For evaluation of (4.15) we use the expansion formula ((2.27) in [18], p.678)
$$
\begin{array}{l}
\displaystyle  A_2^{\left( 3 \right)} \left( {a;b_1 ,b_2 ;c_1 ,c_2 ;x,y,z} \right) \hfill \\
= \displaystyle \left( {1 - x} \right)^{ - b_1 } \left( {1 - y} \right)^{ - b_2 } \sum\limits_{i,j = 0}^\infty  {} \frac{{\left( a \right)_{i - j} \left( {b_1 } \right)_i \left( {b_2 } \right)_i }}
{{\left( {c_1 } \right)_i \left( {c_2 } \right)_i i!j!}}\left( {\frac{x}
{{1 - x}}} \right)^i \left( {\frac{y}
{{1 - y}}} \right)^i z^j  \hfill \\
\times \displaystyle  F\left( {c_1  - a + j,b_1  + i;c_1  + i;\frac{x}
{{x - 1}}} \right)F\left( {c_2  - a + j,b_2  + i;c_2  + i;\frac{y}
{{y - 1}}} \right), \hfill \\
\end{array}\eqno(4.16)
$$
where $F\left( {a,b;c;x} \right)$ is the Gauss hypergeometric function ([7], p. 69, formula (2)). Hence we obtain
$$
\begin{array}{l}
\displaystyle A_2^{\left( 3 \right)} \left( {3 - \alpha  - \beta ;1 - \alpha ,1 - \beta ;2 - 2\alpha ,2 - 2\beta ;\xi ,\eta ,\zeta } \right) \hfill \\
= \displaystyle \left( {\rho ^2 } \right)^{2 - \alpha  - \beta } \left( {\rho ^2  + 4x_0^2  + 4x_0 \rho \cos \,\varphi } \right)^{\alpha  - 1} \left( {\rho ^2  + 4y_0^2  + 4y_0 \rho \sin \,\varphi } \right)^{\beta  - 1} P_{11} , \hfill \\
\end{array}\eqno(4.17)
$$
where
$$
\begin{array}{l}
\displaystyle   P_{11}  = \sum\limits_{i,j = 0}^\infty  {} \frac{{\left( {3 - \alpha  - \beta } \right)_{i - j} \left( {1 - \alpha } \right)_i \left( {1 - \beta } \right)_i }}
{{\left( {2 - 2\alpha } \right)_i \left( {2 - 2\beta } \right)_i i!j!}} \hfill \\
\displaystyle \times \left( {\frac{{4x_0^2  + 4x_0 \rho \cos \,\varphi }}
{{\rho ^2  + 4x_0^2  + 4x_0 \rho \cos \,\varphi }}} \right)^i \left( {\frac{{4y_0^2  + 4y_0 \rho \sin \,\varphi }}
{{\rho ^2  + 4y_0^2  + 4y_0 \rho \sin \,\varphi }}} \right)^i \left( { - \frac{{\lambda ^2 }}
{4}\rho ^2 } \right)^j  \hfill \\
 \displaystyle   \times F\left( { - \alpha  + \beta  - 1 + j,1 - \alpha  + i;2 - 2\alpha  + i;\frac{{4x_0^2  + 4x_0 \rho \cos \,\varphi }}
{{\rho ^2  + 4x_0^2  + 4x_0 \rho \cos \,\varphi }}} \right) \hfill \\
 \times \displaystyle  F\left( {\alpha  - \beta  - 1 + j,1 - \beta  + i;2 - 2\beta  + i;\frac{{4y_0^2  + 4y_0 \rho \sin \,\varphi }}
{{\rho ^2  + 4y_0^2  + 4y_0 \rho \sin \,\varphi }}} \right), \hfill \\
\end{array}\eqno(4.18)
$$
Using equality (46) in ([7], p. 112,)
$$
F\left( {a,b;c;1} \right) = \frac{{\Gamma \left( c \right)\Gamma \left( {c - a - b} \right)}}
{{\Gamma \left( {c - a} \right)\Gamma \left( {c - b} \right)}},\,\,\, c \ne 0, - 1, - 2,...,\,\,Re \left( {c - a - b} \right) > 0,
$$
it is not complicated to calculate
$$
\mathop {\lim }\limits_{\rho  \to 0} P_{11}  = \frac{{\Gamma \left( {2 - 2\alpha } \right)\Gamma \left( {2 - 2\beta } \right)}}
{{\Gamma \left( {1 - \alpha } \right)\Gamma \left( {1 - \beta } \right)\Gamma \left( {3 - \alpha  - \beta } \right)}}.\eqno(4.19)
$$
By virtue of (4.17) we calculate $J_{11}$:
$$
\begin{array}{l}
 \displaystyle J_{11}  = 2k_4 \left( {2 - \alpha  - \beta } \right)x_0^{1 - 2\alpha } y_0^{1 - 2\beta } \int\limits_0^{2\pi } {} \left( {x_0  + \rho \cos \,\varphi } \right)\left( {y_0  + \rho \sin \,\varphi } \right)u\left( {x_0  + \rho \cos \,\varphi ,y_0  + \rho \sin \,\varphi } \right) \hfill \\
\times \left( {\rho ^2  + 4x_0^2  + 4x_0 \rho \cos \,\varphi } \right)^{\alpha  - 1} \left( {\rho ^2  + 4y_0^2  + 4y_0 \rho \sin \,\varphi } \right)^{\beta  - 1} P_{11} d\varphi . \hfill \\
\end{array}
$$
Passing to the limit as $\rho  \to 0^+$ and taking into account (1.8), we have
$$
\mathop {\lim }\limits_{\rho  \to 0} J_{11}  = u\left( {x_0 ,y_0 } \right).\eqno(4.20)
$$
Similarly it can be proved that
$$
\mathop {\lim }\limits_{\rho  \to 0} J_{12}  = \mathop {\lim }\limits_{\rho  \to 0} J_{13}  = \mathop {\lim }\limits_{\rho  \to 0} J_{14}  = \mathop {\lim }\limits_{\rho  \to 0} J_2  = 0, \mathop {\lim }\limits_{\rho  \to 0} \int\limits_{C_\rho  } {x^{2\alpha } y^{2\beta } G_4 \left( {x,y;x_0 ,y_0 } \right)\frac{{\partial u}}
{{\partial n}}ds}  = 0.\eqno(4.21)
$$
Thus, using on equalities (4.13), (4.14), (4.20) and (4.21), by (4.4) we deduce that
$$
\begin{array}{l}
\displaystyle u\left( {x_0 ,y_0 } \right) = k_4 \left( {1 - 2\beta } \right)x_0^{1 - 2\alpha } y_0^{1 - 2\beta }  \hfill \\
\times  \displaystyle \int\limits_0^a {x\tau _1 \left( x \right)\left\{ {\frac{{H_3 \left( {2 - \alpha  - \beta ,1 - \alpha ;2 - 2\alpha ;\rho _1 ,\rho _1^* } \right)}}
{{\left[ {\left( {x - x_0 } \right)^2  + y_0^2 } \right]^{2 - \alpha  - \beta } }} - \frac{{H_3 \left( {2 - \alpha  - \beta ,1 - \alpha ;2 - 2\alpha ;\rho _2 ,\rho _2^* } \right)}}
{{\left[ {\left( {a - \displaystyle  \frac{{xx_0 }}
{a}} \right)^2  +  \displaystyle \frac{1}
{{a^2 }}x^2 y_0^2 } \right]^{2 - \alpha  - \beta } }}} \right\}dx}  \hfill \\
+ k_4 \left( {1 - 2\alpha } \right)x_0^{1 - 2\alpha } y_0^{1 - 2\beta }  \hfill \\
\times  \displaystyle \int\limits_0^a {y\tau _2 \left( y \right)\left\{ {\frac{{H_3 \left( {2 - \alpha  - \beta ,1 - \beta ;2 - 2\beta ;\rho _3 ,\rho _3^* } \right)}}
{{\left[ {x_0^2  + \left( {y - y_0^2 } \right)^2 } \right]^{2 - \alpha  - \beta } }} - \frac{{H_3 \left( {2 - \alpha  - \beta ,1 - \beta ;2 - 2\beta ;\rho _4 ,\rho _4^* } \right)}}
{{\left[ {\left( {a -  \displaystyle \frac{{yy_0 }}
{a}} \right)^2  +  \displaystyle \frac{1}
{{a^2 }}x_0^2 y^2 } \right]^{2 - \alpha  - \beta } }}} \right\}dy}  \hfill \\
-  \displaystyle \int\limits_\Gamma  {x^{2\alpha } y^{2\beta } \varphi \left( s \right)\frac{{\partial G_4 \left( {x,y;x_0 ,y_0 } \right)}}
{{\partial n}}ds} . \hfill \\
\end{array}\eqno(4.22)
$$
If we use the formula
$$
H_3 \left( {a,b;c;x,y} \right) = \left( {1 - x} \right)^{ - b} F_{0:1;2}^{1:1;0} \left[ {\begin{array}{*{20}c}
   {c - a:}  \\
   {\,\,\,\,\,\,\, - :}  \\
 \end{array} \begin{array}{*{20}c}
   {b;}  \\
   {c;}  \\
 \end{array} \begin{array}{*{20}c}
   {\,\,\,\,\,\,\,\,\,\,\,\,\,\,\,\,\,\, - ;}  \\
   {1 - a,c - a;}  \\
 \end{array} \frac{x}
{{x - 1}}, - y} \right],
$$
which connects Kummer's function with the hypergeometric function of Kampe de Feriet ([4], p. 150, formula (29))
$$
\displaystyle F_{\displaystyle l:m;n;}^{\displaystyle p:q;k;} \left[ {\begin{array}{*{20}c}
   {\left( {a_p } \right):}  \\
   {\left( {\alpha _l } \right):}  \\
 \end{array} \begin{array}{*{20}c}
   {\left( {b_q } \right);}  \\
   {\left( {\beta _m } \right);}  \\
 \end{array} \begin{array}{*{20}c}
   {\left( {c_k } \right);}  \\
   {\left( {\gamma _n } \right);}  \\
 \end{array} x,y} \right] = \displaystyle \sum\limits_{ r,s = 0}^\infty  {} \frac{{\prod\limits_{j = 1}^p {} \left( {a_j } \right)_{r + s} \prod\limits_{j = 1}^q {} \left( {b_j } \right)_r \prod\limits_{j = 1}^k {} \left( {c_j } \right)_s }}
{{\prod\limits_{j = 1}^l {} \left( {\alpha _j } \right)_{r + s} \prod\limits_{j = 1}^m {} \left( {\beta _j } \right)_r \prod\limits_{j = 1}^n {} \left( {\gamma _j } \right)_s r!s!}}\displaystyle x^r y^s ,
$$
we find that solution (4.22) of the problem $D$ may be represented as
$$
\begin{array}{l}
\displaystyle u\left( {x_0 ,y_0 } \right) = k_4 \left( {1 - 2\beta } \right)x_0^{1 - 2\alpha } y_0^{1 - 2\beta } \int\limits_0^a {} x\tau _1 \left( x \right)\left\{ {\frac{{F_{0:1;2}^{1:1;0} \left[ {\begin{array}{*{20}c}
   {\beta  - \alpha :}  \\
   {\,\,\,\,\,\,\,\,\ - :}  \\
 \end{array} \begin{array}{*{20}c}
   {\,\,\,1 - \alpha ;}  \\
   {2 - 2\alpha ;}  \\
 \end{array} \begin{array}{*{20}c}
   {\,\,\,\,\,\,\,\,\,\,\,\,\,\,\,\,\,\,\,\,\,\,\,\,\,\,\,\,\,\,\,\,\,\,\,\,\,\ - ;}  \\
   {\alpha  + \beta  - 1,\beta  - \alpha ;}  \\
 \end{array} \sigma _1 ,\sigma _1^* } \right]}}
{{\left[ {\left( {x - x_0 } \right)^2  + y_0^2 } \right]^{1 - \beta } \left[ {\left( {x + x_0 } \right)^2  + y_0^2 } \right]^{1 - \alpha } }}} \right.\, \hfill \\
  \, - \displaystyle  \left. {\frac{{F_{0:1;2}^{1:1;0} \left[ {\begin{array}{*{20}c}
   {\beta  - \alpha :}  \\
   {\,\,\,\,\,\,\,\,\ - :}  \\
 \end{array} \begin{array}{*{20}c}
   {\,\,\,1 - \alpha ;}  \\
   {2 - 2\alpha ;}  \\
 \end{array} \begin{array}{*{20}c}
   {\,\,\,\,\,\,\,\,\,\,\,\,\,\,\,\,\,\,\,\,\,\,\,\,\,\,\,\,\,\,\,\,\,\,\,\ - ;}  \\
   {\alpha  + \beta  - 1,\beta  - \alpha ;}  \\
 \end{array} \sigma _2 ,\sigma _2^* } \right]}}
{{\left[ {\left( {a - \displaystyle \frac{{xx_0 }}
{a}} \right)^2  + \displaystyle \frac{1}
{{a^2 }}x^2 y_0^2 } \right]^{1 - \beta } \left[ {\left( {a + \displaystyle \frac{{xx_0 }}
{a}} \right)^2  + \displaystyle \frac{1}
{{a^2 }}x^2 y_0^2 } \right]^{1 - \alpha } }}} \right\}dx \hfill \\
\end{array}
$$
$$
\begin{array}{l}
   + k_4 \left( {1 - 2\alpha } \right)x_0^{1 - 2\alpha } y_0^{1 - 2\beta } \displaystyle \int\limits_0^a {} y\tau _2 \left( y \right)\left\{ {\displaystyle \frac{{F_{0:1;2}^{1:1;0} \left[ {\begin{array}{*{20}c}
   {\alpha  - \beta :}  \\
   {\,\,\,\,\,\,\,\,\ - :}  \\
 \end{array} \begin{array}{*{20}c}
   {\,\,\,1 - \beta ;}  \\
   {2 - 2\beta ;}  \\
 \end{array} \begin{array}{*{20}c}
   {\,\,\,\,\,\,\,\,\,\,\,\,\,\,\,\,\,\,\,\,\,\,\,\,\,\,\,\,\,\,\,\,\,\,\,\,\ - ;}  \\
   {\alpha  + \beta  - 1,\alpha  - \beta ;}  \\
 \end{array} \sigma _3 ,\sigma _3^* } \right]}}
{{\left[ {x_0^2  + \left( {y - y_0^2 } \right)^2 } \right]^{1 - \alpha } \left[ {x_0^2  + \left( {y + y_0 } \right)^2 } \right]^{1 - \beta } }}} \right. \hfill \\
   - \left. {\displaystyle \frac{{F_{0:1;2}^{1:1;0} \left[ {\begin{array}{*{20}c}
   {\alpha  - \beta :}  \\
   {\,\,\,\,\,\,\,\,\ - :}  \\
 \end{array} \begin{array}{*{20}c}
   {\,\,\,1 - \beta ;}  \\
   {2 - 2\beta ;}  \\
 \end{array} \begin{array}{*{20}c}
   {\,\,\,\,\,\,\,\,\,\,\,\,\,\,\,\,\,\,\,\,\,\,\,\,\,\,\,\,\,\,\,\,\,\,\,\,\ - ;}  \\
   {\alpha  + \beta  - 1,\alpha  - \beta ;}  \\
 \end{array} \sigma _4 ,\sigma _4^* } \right]}}
{{\left[ {\left( \displaystyle {a - \displaystyle \frac{{yy_0 }}
{a}} \right)^2  + \displaystyle \frac{1}
{{a^2 }}x_0^2 y^2 } \right]^{1 - \alpha } \left[ {\left( \displaystyle {a + \displaystyle \frac{{yy_0 }}
{a}} \right)^2  + \displaystyle \frac{1}
{{a^2 }}x_0^2 y^2 } \right]^{1 - \beta } }}} \right\}dy \\
- \displaystyle \int\limits_\Gamma  {x^{2\alpha } y^{2\beta } \varphi \left( s \right)\frac{{\partial G_4 \left( {x,y;x_0 ,y_0 } \right)}}
{{\partial n}}ds,}  \hfill \\
\end{array}\eqno(4.23)
$$
where
$$
\sigma _1  = \frac{{4xx_0 }}
{{\left( {x + x_0 } \right)^2  + y_0^2 }}, \,\,\,\,\,\,\ \sigma _1^*  = \frac{{\lambda ^2 }}
{4}\left[ {\left( {x - x_0 } \right)^2  + y_0^2 } \right],
$$
$$
\sigma _2  = \displaystyle \frac{{4xx_0 }}
{{\left( \displaystyle {a + \frac{{xx_0 }}
{a}} \right)^2  + \displaystyle \frac{1}
{{a^2 }}x^2 y_0^2 }},\,\,\,\,\ \sigma _2^*  = \displaystyle \frac{{a^2 \lambda ^2 }}
{{4R_0^2 }}\left[ {\left( \displaystyle {a - \frac{{xx_0 }}
{a}} \right)^2  + \displaystyle \frac{1}
{{a^2 }}x^2 y_0^2 } \right],
$$
$$
\sigma _3  = \frac{{4yy_0 }}
{{x_0^2  + \left( {y + y_0 } \right)^2 }}, \,\,\,\,\,\,\,\ \sigma _3^*  = \frac{{\lambda ^2 }}
{4}\left[ {x_0^2  + \left( {y - y_0 } \right)^2 } \right],
$$
$$
\sigma _4  = \displaystyle \frac{{4yy_0 }}
{{\left( \displaystyle {a + \frac{{yy_0 }}
{a}} \right)^2  + \displaystyle \frac{1}
{{a^2 }}x_0^2 y^2 }}, \,\,\,\,\,\,\,\,\ \sigma _4^*  = \displaystyle \frac{{a^2 \lambda ^2 }}
{{4R_0^2 }}\left[ {\left(\displaystyle  {a - \frac{{yy_0 }}
{a}} \right)^2  + \displaystyle \frac{1}
{{a^2 }}x_0^2 y^2 } \right].
$$

Now we can formulate the main result.

\textbf{Theorem 2.} The problem $D$ has the unique solution defined by formula (4.23).

	We note that expansions for the hypergeometric functions of Lauricella $F_A^{\left( n \right)} ,\,\,F_B^{\left( n \right)} ,\,\,F_C^{\left( n \right)} ,\,\,F_D^{\left( n \right)}
$ are found in [19, 20] and applied in [17] for finding fundamental solutions and later for investigating boundary value problems for 3-D singular elliptic equations [24].

\section{Acknowledgement}. We would like to thank the anonymous referee for valuable suggestions, which made the present paper more readable.

\bigskip
Makhmud Salakhitdinovich Salakhitdinov\\
Institute of Mathematics, National University of Uzbekistan \\
29 Durmon yuli st., Tashkent 100125, Uzbekistan \\
E-Mail: salakhitdinovms@yahoo.com, \\
\medskip
Anvar Hasanov\\
Institute of Mathematics, National University of Uzbekistan \\
29 Durmon yuli st., Tashkent 100125, Uzbekistan \\
E-Mail: anvarhasanov@yahoo.com\\
\medskip
\begin{flushright}
Received: 28.09.2012
\end{flushright}


\begin{thebibliography}{99}

\bibitem{hb 1}
A.Altin ,  \emph{Solutions of type $r^m $ for a class of singular equations}. International Journal of Mathematical Science, 5(3)(1982), 613-619.

\bibitem{hb 2}
A.Altin, \emph{Some expansion formulas for a class of singular partial differential equations}. Proceedings of American Mathematical Society, 85(1) (1982), 42-46.

\bibitem{hb 3}
A.Altin, Y.Eutiquio, \emph{Some properties of solutions of a class of singular partial differential equations}. Bulletin of the Institute of Mathematics Academic Sinica, 11(1) (1983), 81-87.

\bibitem{hb 4}
P.Appell, J.Kampe de Feriet, \emph{Fonctions Hypergeometriques et Hyperspheriques;  Polynomes d'Hermite}, Gauthier - Villars. Paris, 1926.

\bibitem{hb 5}
A.Erdelyi, \emph{Singularities of generalized axially symmetric potentials}. Comm. Pure Appl. Math., 2 (1956), 403-414.

\bibitem{hb 6}
A.Erdelyi, \emph{An application of fractional integrals}. J. Analyse. Math., 14 (1965), 113-126.

\bibitem{hb 7}
A.Erdelyi, W.Magnus, F.Oberhettinger, F.G.Tricomi, \emph{Higher transcendental
functions} (Russian), vol. I, Izdat. Nauka, Moscow, 1973.

\bibitem{hb 8}
A.J.Fryant, \emph{Growth and complete sequences of generalized bi-axially symmetric potentials}. Journal of Differential Equations, 31(2) (1979), 155-164.

\bibitem{hb 9}
R.Gilbert, \emph{On the singularities of generalized axially symmetric potentials}. Arch. Rational Mech. Anal., 6 (1960), 171-176.

\bibitem{hb 10}
R.Gilbert, \emph{Some properties of generalized axially symmetric potentials}. Amer. J. Math., 84 (1962), 475-484.

\bibitem{hb 11}
R.Gilbert, \emph{"Bergman's" integral operator method in generalized axially symmetric potential theory}. J. Mathematical Phys., 5 (1964), 983-987.

\bibitem{hb 12}
R.Gilbert, \emph{On the location of singularities of a class of elliptic partial differential equations in four variables}. Canad. J. Math., 17 (1965), 676-686.

\bibitem{hb 13}
R.Gilbert, H.Howard, \emph{On solutions of the generalized axially symmetric wave equation represented by Bergman operators}, Proc. London Math. Soc., 15 (2) (1965), 346-360.

\bibitem{hb 14}
R.Gilbert, \emph{On the analytic properties of solutions to a generalized axially symmetric Schroedinger equations}. J. Differential equations, 3 (1967), 59-77.

\bibitem{hb 15}
R.Gilbert, \emph{An investigation of the analytic properties of solutions to the generalized axially symmetric, reduced wave equation in $n + 1$  variables, with an application to the theory of potential scattering}. SIAM J. Appl. Math. 16 (1) (1968), 30-50.

\bibitem{hb 16}
R.Gilbert, \emph{Function Theoretic Methods in Partial Differential Equations} (New York, London: Academic Press), 1969.

\bibitem{hb 17}
A.Hasanov, E.T.Karimov, \emph{Fundamental solutions for a class of three-dimensional elliptic equations with singular coefficients}. Applied Mathematics Letters, 22 (2009), pp. 1828-1832.

\bibitem{hb 18}
A.Hasanov, \emph{Fundamental solutions of generalized bi-axially symmetric Helmholtz equation}. Complex Variables and Elliptic Equations, 52(8) (2007), 673-683.

\bibitem{hb 19}
A.Hasanov, H.M.Srivastava, \emph{Decomposition Formulas Associated with the Lauricella Multivariable Hypergeometric Functions}. Computers and Mathematics with Applications, 53(7) (2007), 1119-1128.

\bibitem{hb 20}
A.Hasanov, H.M.Srivastava, \emph{Some decomposition formulas associated with the Lauricella Function $F_A^{\left( r \right)}$ and other multiple hypergeometric functions}. Applied Mathematics Letters, 19(2) (2006), 113-121.

\bibitem{hb 21}
P.Henrici, \emph{On the domain of regularity of generalized axially symmetric potentials}. Proc. Amer. Math. Soc., 8 (1957), 29-31.

\bibitem{hb 22}
P.Henrici, \emph{Complete systems of solutions for a class of singular elliptic Partial Differential Equations. Boundary Value Problems in differential equations}, University of Wisconsin Press, Madison, (1960), 19-34.

\bibitem{hb 23}
A.Huber, \emph{On the uniqueness of generalized axisymmetric potentials}. Ann. Math., 60 (1954), 351-358.

\bibitem{hb 24}
E.T.Karimov, \emph{On a boundary problem with Neumann's condition for 3-D singular
elliptic equations}. Applied Mathematics Letters, 23 (2010), 517-522.

\bibitem{hb 25}
D.Kumar, \emph{Approximation of growth numbers generalized bi-axially symmetric potentials}. Fasciculi Mathematics, 35 (2005), 51-60.

\bibitem{hb 26}
C.Y.Lo, \emph{Boundary value problems of generalized axially symmetric Helmholtz equations}. Portugaliae Mathematica, 36(3-4) (1977), 279-289.

\bibitem{hb 27}
O.I.Marichev, \emph{Integral representation of solutions of the generalized double axial symmetric Helmholtz equation} (Russian). Differencial'nye Uravnenija, Minsk, 14(10) (1978), 1824-1831.

\bibitem{hb 28}
P.A.McCoy, \emph{Polynomial approximation and growth of generalized axisymmetric potentials}. Canadian Journal of Mathematics, 31(1) (1979), 49-59.

\bibitem{hb 29}
K.B.Ranger, \emph{On the construction of some integral operators for generalized axially symmetric harmonic and stream functions}. J. Math. Mech., 14, (1965) 383-402.

\bibitem{hb 30}
J.M.Rassias, A.Hasanov, \emph{Fundamental Solutions of Two Degenerated Elliptic Equations and Solutions of Boundary Value Problems in Infinite Area}. International Journal of Applied Mathematics and Statistics, 8(7) (2007), 87-95.

\bibitem{hb 31}
M.S.Salakhitdinov, A.Hasanov, \emph{A solution of the Neumann-Dirichlet boundary value problem for generalized bi-axially symmetric Helmholtz equation}. Complex Variables and Elliptic Equations, 53 (4) (2008), 355-364.

\bibitem{hb 32}
R.J.Weinacht, \emph{Some properties of generalized axially symmetric Helmholtz potentials}. SIAM J. Math. Anal. 5 (1974), 147-152.

\bibitem{hb 33}
A.Weinstein, \emph{Discontinuous integrals and generalized potential theory}. Trans. Amer. Math. Soc., 63 (1948), 342-354.

\bibitem{hb 34}
A.Weinstein, \emph{Generalized axially symmetric potentials theory}. Bull. Amer. Math. Soc., 59 (1952), 20-38.

\end{thebibliography}
\end{document}